\documentclass[11pt]{article}

\usepackage[letterpaper]{geometry}

\usepackage[utf8]{inputenc} 
\usepackage[T1]{fontenc}    
\usepackage[semibold,tabular]{sourceserifpro}
\usepackage[erewhon,scaled=1.03]{newtxmath}
\usepackage[scaled=1.1]{zlmtt}
\usepackage{amsfonts}
\usepackage[final,letterspace=25,nopatch=footnote]{microtype}
\usepackage{ragged2e}
\usepackage{siunitx}

\usepackage[onehalfspacing]{setspace}
\usepackage{xspace}

\usepackage[english]{babel}

\usepackage{amsmath}
\usepackage[noblocks]{authblk}
\usepackage{booktabs}
\usepackage{graphicx}
\usepackage[colorlinks=true, allcolors=blue]{hyperref}

\usepackage{natbib}

\newcommand{\CAPS}[2]{\textls[50]{\textsc{\MakeLowercase{#1}}}\textls[50]{#2}}
\newcommand{\caps}[1]{\CAPS{#1}{}}

\newcommand{\PySAGES}{\CAPS{P}{y}\caps{SAGES}\xspace}

\title{
  Autodifferentiable Geometric Restraints for Enhanced Sampling Simulations with Classical and Machine Learned Force Fields
}

\author[1]{\small Gustavo R. Pérez-Lemus}
\author[1]{Cintia A. Menendez}
\author[1]{Yinan Xu}
\author[1]{Pablo F. {Zubieta~Rico}}
\author[1]{Yezhi Jin}
\author[2]{Juan J. {de~Pablo}}

\affil[1]{\footnotesize Pritzker School of Molecular Engineering, The University of Chicago, Chicago, IL 60615, USA}
\affil[2]{Materials Science Division, Argonne National Laboratory, Argonne, IL 60439, USA}

\date{}

\begin{document}

\maketitle

\begin{abstract}
The use of external restraints is ubiquitous in advanced molecular simulation techniques. In general, restraints serve to reduce the configurational space that is available for sampling, thereby reducing the computational demands associated with a given simulations. Examples include the use of positional restraints in docking simulations or positional restraints in studies of catalysis. Past work has sought to couple complex restraining potentials with enhanced sampling methods, including Metadynamics or Extended Adaptive Biasing Force approaches. Here, we introduce the use of more general geometric potentials coupled with enhanced sampling methods that incorporate neural networks or spectral decomposition to achieve more efficient sampling in the context of advanced materials design.
\end{abstract}

\section{Introduction}
Simulation of the free-energy landscapes that govern complex molecular transformations is a central element of statistical and quantum mechanics. This is the case for biological processes such as protein-ligand binding~\cite{prot-lig}, chemical processes such as heterogeneous catalysis~\cite{piccini2022ab}, and physical processes such as reorientation of liquid crystals at interfaces~\cite{ramezani2017understanding}. These systems are commonly modeled using molecular models and molecular dynamics simulations, where atoms within the system interact through potential energies calculated from, for instance, density functional theory (\caps{DFT})~\cite{burke2013dft}, classical models~\cite{brooks2021classical}, or, recently, machine-learned force fields (\caps{MLFF})~\cite{unke2021machine}. However, accurate calculation of free energies via molecular dynamics often requires the use of enhanced sampling methods. This requirement is due to the significant energy barriers that separate various states in many processes of interest, which, in many cases, cannot be overcome merely by the thermal motion of the particles.

Enhanced sampling methods are essential for molecular dynamics simulations of complex systems as they facilitate the exploration of extensive portions of the configuration space and enable the calculation of free-energy profiles.
Given the large family of enhanced sampling methods~\cite{henin2022enhanced}, in this work, we focus on those that build a bias potential to later calculate the free energy using a collective variable in the system.
Examples of these methods include Metadynamics~\cite{wtmetad} and its new upgrade \caps{OPES}~\cite{invernizzi2020unified}, Umbrella Sampling~\cite{torrie1977nonphysical}, Adaptive Biasing Force (\caps{ABF})~\cite{comer2015adaptive}, and the recently-proposed Extended Adaptive Biasing Force (e\caps{ABF})~\cite{fu2016extended}.
These methods have been successfully applied to a wide variety of biological processes~\cite{decherchi2020thermodynamics}.
A key application is the accurate determination of the binding free energies of potential drugs against target proteins.
In this context, the enhanced sampling method is used to calculate the potential of mean force (\caps{PMF}) to remove the ligand from its binding site. On the other hand, to address potential sampling challenges in \caps{PMF} calculations, it is essential to use external restraints that limit the configurational space and promote convergence in the \caps{PMF} calculation. In addition, it is necessary to analyze the effect of the restraining potential in the calculation and adjust for its influence to accurately report binding energies, which are critical for differentiating between drug candidates. For these purposes, protocols such as \caps{BFEE}~\cite{fu2022accurate,woo2005calculation} have integrated e\caps{ABF} with a set of positional and orientational restraints to calculate the \caps{PMF} along a defined spatial path of a ligand. Funnel Metadynamics is another powerful enhanced sampling method that has been used to study a wide variety of systems~\cite{limongelli2013funnel}. This method applies a funnel-shaped spatial restraint on a grid to control the movement of a ligand as it is detached from the binding pose following a determined path. However, methods such as \caps{ABF} have not yet incorporated external restraints. Recent advancements in the \caps{ABF} family of methods, such as the incorporation of neural networks or spectral decomposition, have shown to markedly enhance the speed of convergence in free energy calculations. The integration of spatial restraints in these enhanced methods can lead to even greater efficiency and accuracy in their application.

Such integration of spatial restraints to the \caps{ABF} family enhanced sampling is especially relevant given the recent advancements in Machine Learned Force Fields (\caps{MLFF}s). \caps{MLFF}s are surging as a new paradigm for molecular dynamics simulations~\cite{chmiela2023accurate}. These \caps{MLFF}s rely on training a neural network from a dataset of atomic configurations labeled with energies and forces. From that dataset, the neural network can learn the potential energy surface of the system based on local descriptors that can be invariant~\cite{deepmd} or equivariant~\cite{allegro}. The growing interest in \caps{MLFF} is driven by their ability to maintain \caps{DFT}-level accuracy in energy and force predictions at a fraction of the computational cost. Moreover, the expansive capability of these models makes them well-suited for complex scenarios such as catalysis, which often involve bond formation and breaking -- tasks that are challenging for traditional classical force fields that rely on fixed system topologies~\cite{gissinger2017modeling}. For this latter application, \caps{MLFF} has been successfully applied in heterogenous catalysis~\cite{nitrogen}. These modeled systems share features with biological systems like protein-ligand binding, where there is an active binding site and a complex detaching path that the reactants can follow.  In such systems, employing geometrical restraints can be particularly beneficial, accelerating the calculation of free energies when combined with enhanced sampling techniques and computational tools like just-in-time compilation and automatic differentiation (\caps{AD}).

In this work we are proposing the incorporation of a general geometric restraints in the family of \caps{ABF} methods implemented in the \PySAGES enhanced sampling library that facilitated the application of the restraints using \caps{AD} and \caps{GPU} support~\cite{zubieta2024pysages}. We show how this methods works in classical systems and in \caps{MLFF} applications. Additionally, we incorporate Funnel Metadynamics into \PySAGES with the advantage of grid-free restraint potentials that reduces computation cost and \caps{AD} facilitates the incorporation of more complex funnel shape potentials for general application. We hope this work serves the scientific community on applications of free energy methods where the use of restraints are imperative for an accurate calculation of free energies.

\section{Results and Discussion}
In order to incorporate restraint potentials into the \caps{ABF} family of methods, like \caps{ABF}, \caps{FUNN}~\cite{guo2018adaptive}, \caps{CFF}~\cite{sevgen2020combined}, and Spectral \caps{ABF}~\cite{rico2025}, we need to show how we can recover the contribution of the restraint potential on the free energy calculation. For this goal, we use the Darve et al. equation~\cite{darve2008adaptive}:
\begin{equation}
 \nabla_{\xi}A = 
   -\left\langle\frac{d}{dt}(w\cdot p)\right\rangle_{\xi}
\end{equation}
where $\nabla_{\xi}A$ is the gradient of the free energy with respect a collective variable $\xi$, $p$ is the momenta of the particles of the system and $w=M_{\xi}J_{\xi}M^{-1}$ is the mass matrix of the jacobian transformation of $\xi$. Then, Darve et al. show that the time derivative of the $w\cdot p$ matrix is related to the original \caps{ABF} equation:
\begin{equation}
  \left\langle\frac{d}{dt}(w\cdot p)\right\rangle_{\xi} =
    \left\langle-\nabla U\cdot w+\nabla\cdot w\right\rangle_{\xi}
\end{equation}
where $U$ is the potential energy of the system. Using the last equation, we can calculate the contribution of an external restraint applied to the system. Suppose now, that a restraint potential $U_{R}$ is applied to a system, then, the new potential energy of the system is $U'=U+U_{R}$, so the last equation transforms to:
\begin{eqnarray}
     \nabla_{\xi}A'&=&\langle-\nabla (U+U_{R})\cdot w+\nabla\cdot w\rangle_{\xi}\\
     &=&\langle-\nabla U\cdot w+\nabla\cdot w\rangle_{\xi}-\langle\nabla U_{R}\cdot w\rangle_{\xi}\\
     \nabla_{\xi}A&=&\nabla_{\xi}A'+\langle\nabla U_{R}\cdot w\rangle_{\xi}\label{eqfin}
\end{eqnarray}
where eq. \ref{eqfin} give us the way to calculate the correction of the free energy $\nabla_{\xi} A$ when a external potential is applied, since the factor $\langle\nabla U_{R}\cdot w\rangle_{\xi}$ can be stored in similar way as the thermodynamic force $\nabla_{\xi}A'$ in the simulation. Given the generality of this equation, the external potential $U_{R}$ can be any differentiable function of the coordinates and the collective variable $\xi$ can be also very general and include more than one \caps{CV} per calculation. In this case, the use of \caps{AD} in \PySAGES facilitates the incorporation of a wide variety of combinations between external potentials and collective variables.
\subsection{ Recover the correct Free Energy: \caps{ADP} in solution}
\begin{figure}
\centering
\includegraphics[width=1.0\linewidth]{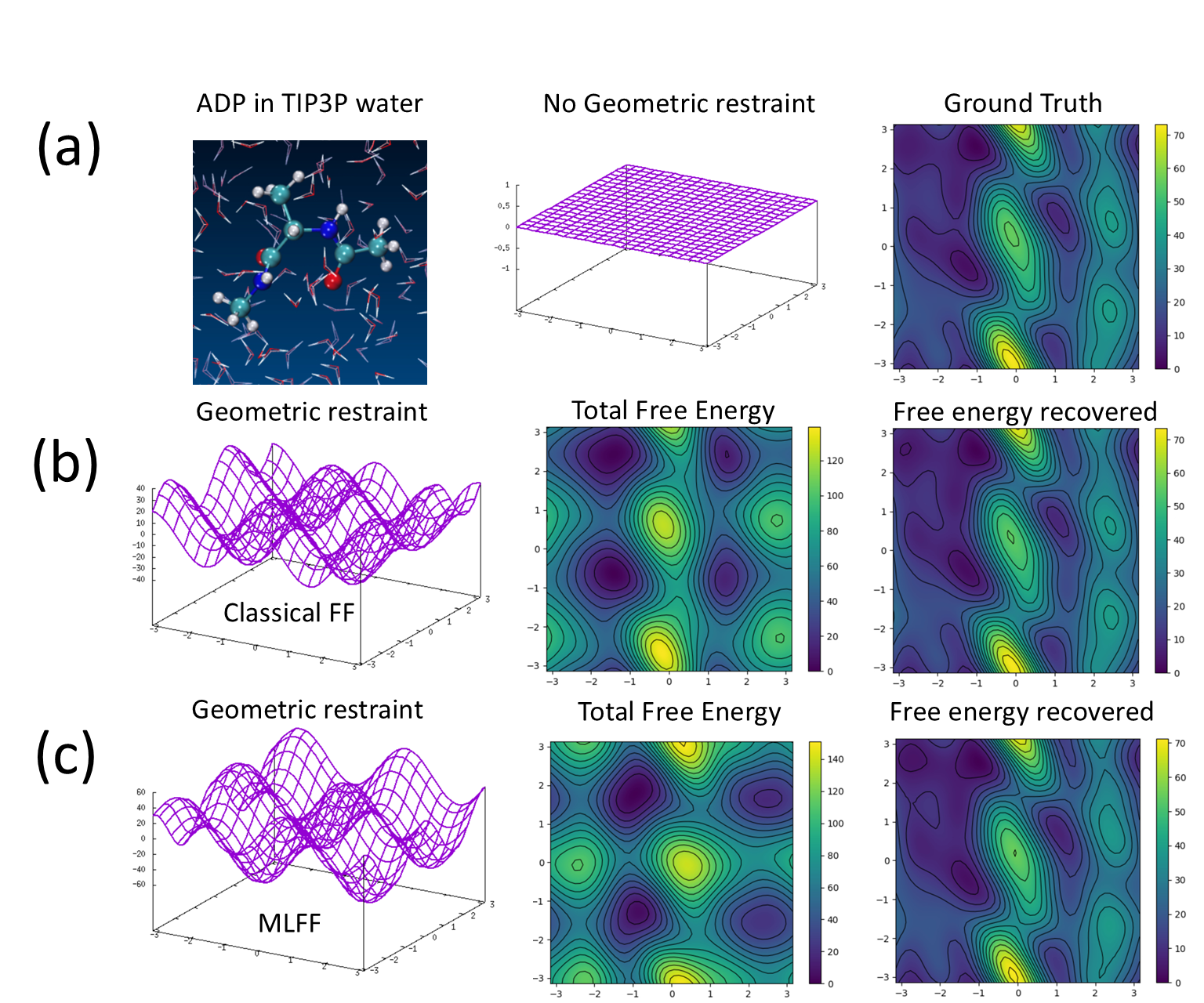}
\caption{\label{fig:adp}Geometric restraint application example. a) Alanine dipeptide in water is modelled with a null restraint applied so we can get the \caps{FES} of the system. A geometric restraint is applied on the b) classical and c) Coarse-grained \caps{MLFF} system, and then the total \caps{FES} is now a combination of the restraint and the underlying \caps{FES}. The eq. \ref{eqfin} allows to recover the correct \caps{FES} for both cases. In all cases the Spectral \caps{ABF} method was used.}
\end{figure}

As first example of the application of eq. \ref{eqfin}, we use as benchmark the solvated alanine dipeptide system (\caps{ADP}), where we add restraints in the shape of $U_{R}(\phi,\psi)=A(cos(b\psi)+sin(c\phi))$ into the system, with $\phi$ and $\psi$ the principal dihedrals of the \caps{ADP} molecule. We use a classical explicit solvated \caps{ADP} molecule in water, and a \caps{MLFF} of \caps{ADP} trained using the forces of the classical simulation, in that way, we have a coarse grained implicit \caps{ADP} \caps{MLFF}~\cite{Perez-Lemus-data}. We use Allegro~\cite{allegro} as the framework to train the model. The results shown in fig. \ref{fig:adp} shows that when the restraint is applied, the total free energy $A'$, shown in the middle column of fig. \ref{fig:adp}, have deviated a lot from the ground truth due to the restraint, however, using the correction given in eq. \ref{eqfin}, we can properly recover $A$ in both classical (fig. \ref{fig:adp}b) and implicit solvent \caps{MLFF} of \caps{ADP} (fig. \ref{fig:adp} c).

\subsection{Accelerate convergence: Host guest \caps{CB8--G6}}
\begin{figure}
\centering
\includegraphics[width=1.0\linewidth]{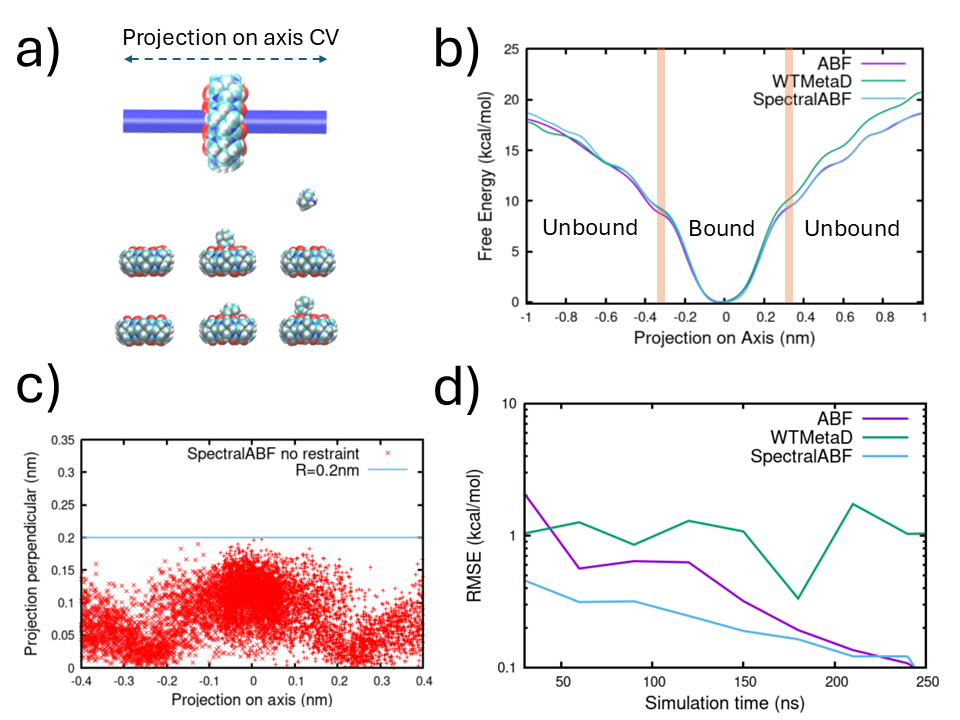}
\caption{\label{fig:cb8} Geometric restraint simulations for the \caps{CB8--G6} system. a) Scheme of the geometric restraint and the cylindrical restrain used in the simulation with snapshot of the system capturing different rebinding events with the geometric restraint in \PySAGES. b) Free energy profiles along the \caps{CV} used at \SI{300}{ns} of simulation, the bound and unbound interface regions are highlighted by orange lines. c) Unrestrained simulation with Spectral \caps{ABF} to select the radius of the cylinder. The radius of \SI{0.2}{nm} allows to sample the system without perturbing the bound mode of the \caps{G6} molecule inside \caps{CB8}. d) Root Mean Square Error (\caps{RMSE}) as a function of the simulation time for the methods applied in the system. Spectral \caps{ABF} has the fastest convergence for all the method studied. Atomic configurations were rendered using the \caps{VMD} software~\cite{HUMP96}.}
\end{figure}

Continuing with the benchmark examples, we have included one of the host molecules of the \caps{SAMPL6} test: the cucurbituril \caps{CB8}~\cite{CB8} along with the guest ligand cycloheptanamide \caps{G6}~\cite{rizzi2018overview}. This molecule have been recently studied using Well-tempered Metadynamics incorporating machine-learning collective variables~\cite{bertazzo2021machine}. For this section, we used a general funnel shape restraint:
\begin{equation}
  U_{R} = \begin{cases}
    \,\dfrac{k}{2} \left(
      \xi_{\perp} -f(\xi_{\parallel})
    \right) ^{2} & \quad \text{if} \quad \xi_{\perp} \geq f(\xi_{\parallel})\\[1em]
    \,0 & \quad \text{otherwise}
  \end{cases}
\end{equation}
where $\xi_{\parallel}$ is the collective variable that measures the projection of the center of mass of the ligand respect to an axis, like in Funnel Metadynamics. $\xi_{\perp}$ is the perpedicular projection to that axis and $f(\xi_{\parallel})$ could be any positive piecewise differentiable function. 
For simplicity, the function $f$ used in this example is $f(\xi)=R_\text{cyl}$, with $R_\text{cyl}=\SI{0.2}{nm}$ as seen in fig. \ref{fig:cb8} a) with snapshots capturing the rebinding events. The radius is chosen to avoid affecting the binding region, as required in the discussion from the use of cylindrical restraints~\cite{roux2008comment}, this was observed in runs in Spectral \caps{ABF} without any restraint (fig. \ref{fig:cb8} c) to capture the unbinding path of the \caps{G6} molecule as it detached from the \caps{CB8} host. After choosing the restraint, then we can apply the same equation as used in~\cite{bertazzo2021machine} to calculate the binding constant~\cite{doudou2009standard}:
\begin{eqnarray}
    \Delta G^{0}_\text{bound}&=&\Delta G_\text{bound}-\Delta G_{v}\\ \label{eq:bind}
    &=&-kT\log\left(\frac{Q_{site}}{Q_{bulk}}\right)-kT\log\left(\frac{V_{bulk}}{V_{0}}\right)\\
    \frac{Q_{site}}{Q_{bulk}}&=&\frac{\int_{site}\exp\left( -\frac{F(\xi)}{kT}\right)d\xi}{\int_{bulk}\exp\left( -\frac{F(\xi)}{kT}\right)d\xi}
\end{eqnarray}
where $\xi$ is the collective variable for the funnel shape, $F(\xi)$ the free energy profile, $V_{0}$ is the reference volume ($1.660\text{\AA}^{3}$), and $V_{bulk}=\int_{bulk}\pi f^{2}(\xi)d\xi$ is the exploration volume that can be approximated by the function $f$ included in the funnel restraint $U_{R}$. The site/bulk regions were chosen as in~\cite{bertazzo2021machine} using the criteria of the first inflection point of $\Delta G$. The binding energy can be calculated using eq. \ref{eq:bind}. The obtained free energy profiles are shown in fig. \ref{fig:cb8} b), where all methods shown similar shape in the bound region and only small deviation in the unbound one. Additionally the Root Mean Square Error (\caps{RMSE}) of the free energy profiles as function of the simulation time are shown in fig. \ref{fig:cb8} d), with Spectral \caps{ABF} reaching 1 kcal/mol convergence in less than 50ns. The calculated binding are shown in table \ref{tab:cb8}, where all the methods studied in this section approximate the binding free energy below the 1kcal/mol range as in~\cite{bertazzo2021machine} but with $\approx4$ times smaller simulation time for all methods accounting of exploring the double of the \caps{CV} range.

\begin{table}
\centering
\begin{tabular}{lr} \toprule
Enhanced Sampling Method & \caps{ABFE} (\si{kcal/mol}) \\ \midrule
\caps{ABF} & -8.24$\pm$ 0.28 \\
s\caps{ABF} & -8.54$\pm$ 0.27\\
WTMetaD & -8.82$\pm$ 0.27 \\
WTMetaD \caps{MLCV}~\cite{bertazzo2021machine} & -8.90$\pm$0.1\\
Experimental & -8.34 \\ \bottomrule
\end{tabular}
\caption{\label{tab:cb8} Absolute Binding Free Energy of \caps{CB8--G6} system with \SI{300}{ns} simulation time. The value of~\cite{bertazzo2021machine} was obtained from \SI{1}{\us} simulation time and using a machine-learned collective variable (\caps{MLCV}).}
\end{table}

\subsection{ Reducing Collective Variables: Trypsin Benzamidine}

\begin{figure}
\centering
\includegraphics[width=1.0\linewidth]{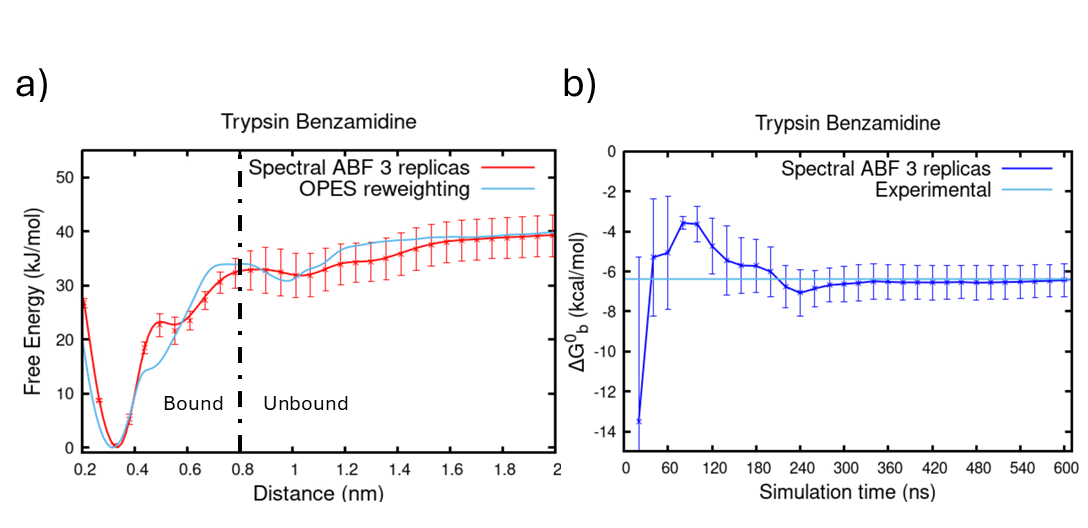}
\caption{\label{fig:prot}Convergence in the free energy landscape of a Ligand-Protein complex.  A) Free Energy curves using Spectral \caps{ABF} with a funnel-like restraint potential. The curve is average over 3 independent replicas. The curve from \caps{OPES} using reweighting over a 2D sampling is included for comparison~\cite{ansari2022water} and the bound/unbound interface as dotted vertical line was obtained from the same reference (\SI{0.8}{nm}). B) Convergence of the $\Delta G_{b}^{0}$ of Benzamidine with Trypsin. The experimental reference is the one used in~\cite{rizzi2023oneopes}.}
\end{figure}

Once we concluded exploratory studies in the \caps{CB8--G6} model system, we explored systems of greater complexity and relevance. The complex Trypsin Benzamidine is a known benchmark system for absolute binding free energy calculations~\cite{ansari2022water}. In this case, following the previous calculation in the literature using external funnel like restraints, the equation used to calculate the binding free energy is:
\begin{equation}
    \Delta G^{0}_\text{bound}=-kT\log \left ( C^{0}\pi R_\text{cyl}^{2}\int_\text{bound} d\xi \exp\left ( -\frac{1}{kT}(F(\xi)-F_\text{unbound})\right ) \right )
\end{equation}
where $C^{0}$ is the standard concentration (\SI{1.660}{\AA^{-3}}), $R_\text{cyl}$ is the cylinder radius in the funnel restraint, $F(\xi)$ is the free energy profile as function of the \caps{CV} $\xi$, and $F_\text{unbound}$ is the free energy value at the unbounded region. Based on the work of One\caps{OPES}~\cite{rizzi2023oneopes}, the interface separating bound/unbound states are set to \SI{0.8}{nm}, and the total restraint applied on this system is:
\begin{equation}
  U_{R} = \begin{cases}
    \,\frac{1}{2}k_{1}(\xi_\text{\caps{RMSD}}-\xi_{0})^{2}+\frac{1}{2}k_{2}\left ( \xi_{\perp} -f(\xi_{\parallel}) \right ) ^{2} & \quad if\quad \xi_{\perp}\geq f(\xi_{\parallel})\\
    \,\frac{1}{2}k_{1}(\xi_\text{\caps{RMSD}}-\xi_{0})^{2} & \quad\text{otherwise}
  \end{cases}
\end{equation}
where $\xi_\text{\caps{RMSD}}$ is the Root Mean Squared Displacement (RMSD) value of the $C_{\alpha}$ carbons of trypsin respect to the crystal structure, $\xi_{0}=0.09$ and $f(\xi_{\parallel})$ takes a logistic shape.
The results in table \ref{tab:bzn} show that Spectral \caps{ABF} is capable of reproducing the experimental result using only a single \caps{CV} (distance) for exploration, as compared to other approaches that require two main \caps{CV}s paired with a set of additional \caps{CV}s~\cite{rizzi2023oneopes}. The free energy curve is shown in fig.\ref{fig:prot} a), where the results shown that for larger distances, the free energy matches the one obtained in~\cite{ansari2022water}. The convergence of the method is shown in fig.\ref{fig:prot} b), where Spectral \caps{ABF} can reach convergence in around 300ns. With these results, we can consider using Spectral \caps{ABF} coupled with geometric restraints as a fast way to calculate absolute binding free energies at the same level as \caps{OPES}.

\begin{table}
\centering
\begin{tabular}{lr} \toprule
Enhanced Sampling Method & \caps{ABFE} (\si{kcal/mol}) \\ \midrule
Spectral \caps{ABF} (1 \caps{CV}) & \num{6.35 \pm 0.87} \\
\caps{OPES} (2 \caps{CV}s) & \num{6.36 \pm 0.07} \\
Experimental & 6.36 \\ \bottomrule
\end{tabular}
\caption{\label{tab:bzn}Absolute protein-ligand binding free energy for Trypsin-Benzamidine complex.}
\end{table}
\subsection{Entropic effects in catalysis: Methane activation on Nickel with \caps{MLFF}}
\begin{figure}
\centering
\includegraphics[width=1.0\linewidth]{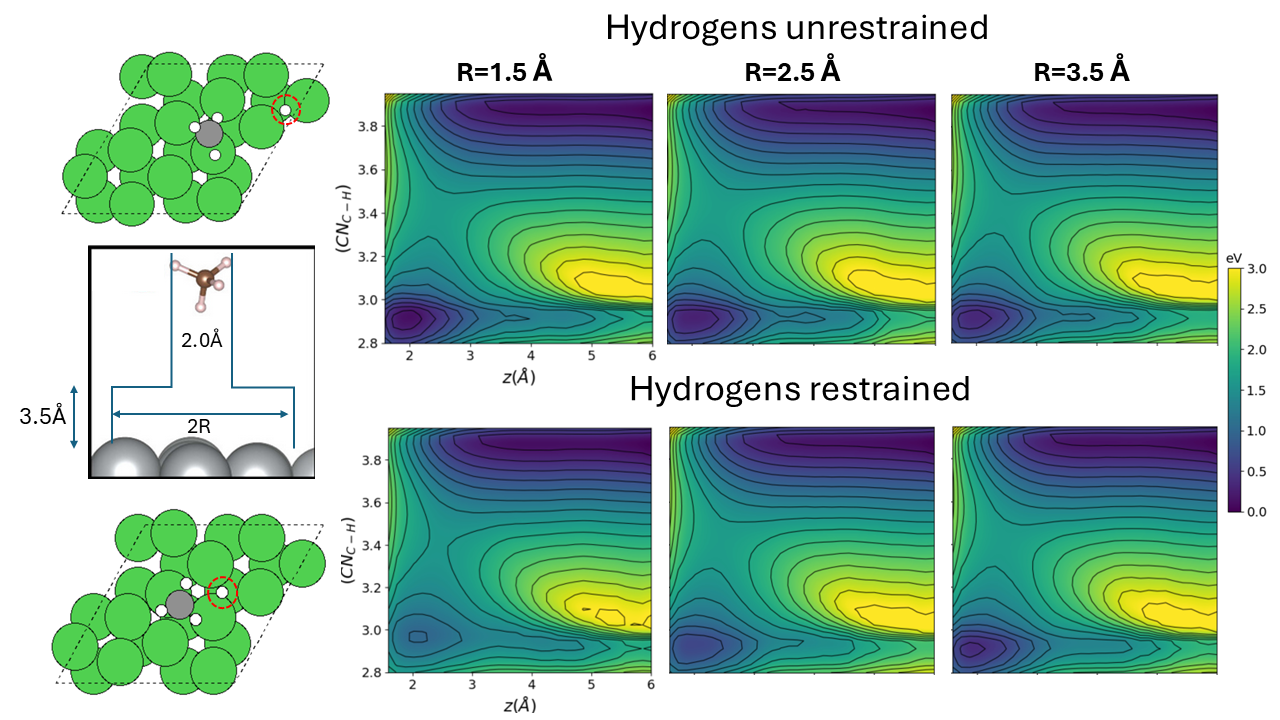}
\caption{\label{fig:n2}External restraints in catalysis simulations for accelerate convergence and entropic contribution. Left: Scheme of the spatial restraint in the system of methane nickel. Right: Free energy profiles of the distance to interface and Coordination number for Carbon-Hydrogen as function of the radius of the restraint ($R$) and the inclusion of hydrogens in the restraint. If hydrogens are not included, the free energy profile is independent of the $R$ parameter, whereas if they are included, the contrary occurs.}
\end{figure}

In this last section, we studied a catalytic process on a metal surface; the methane activation on nickel. Previous work have uncovered the importance of the entropy on the reaction barrier~\cite{xu2024molecular}. Taken this into account, we decide to build a system where we can control the translation of the carbon and methane on the surface to capture the importance of its mobility for the free energy. We build a restraint potential as the one depicted in Figure \ref{fig:n2}. This potential has the shape:
\begin{align}
  U_{R} &= U_\text{Ni}+U_\text{C}+U_\text{H}\\
  U_{i} &= \begin{cases}
    \,\frac{1}{2}k_{1,i}(\mathbf{z}_{i}-z_{i,\max})^{2}+\frac{1}{2}k_{2,i}\left ( r_{i} -f(\mathbf{z}_{i}) \right ) ^{2} & \quad\text{if}\quad r_{i}\geq f(\mathbf{z}_{i})\, or\,\mathbf{z}_{i}\geq z_{i,\max}\\
    \,0 & \quad\text{otherwise} \quad (i=\text{Ni, C, H})
  \end{cases}
\end{align}
where $\mathbf{z}_{i}$ is the z component of the atom $i$ of the system, and $r_{i}$ is the plane component of the position of the atom respect to the same surface. For Ni, $k_{2}=0$ and $z_{\max} = \SI{13.5}{\AA}$ are set to avoid that Ni atoms move too high in $\mathbf{z}$, going outside the training set. For carbon and hydrogen, this restraint can control the diffusion of the system at the surface while it allows to sample faster the gas phase of methane, i.e. sampling distances to the surface larger than \SI{3.5}{\AA}~\cite{xu2024molecular}. The results using \caps{ABF} are displayed the fig. \ref{fig:n2}, which highlight that when the restraint includes hydrogen, the free energy difference between physisorbed methane (CH$_{4}$) and chemisorbed methyl moiety with hydrogen ($\text{CH}^{*}_{3} + \text{H}^{*}$) is dependent on the radius used in the restraint. In contrast, when only the carbon atom is restraint, the \caps{FEP} remains independent of the restraint radius. This is consistent with the previous results~\cite{xu2024molecular}, where the mobility of carbon is restrained by the interaction with nickel, but the hydrogens are free to diffuse on the surface, leading with a significant entropy contribution that a restraint on them could be preventing. 

\section{Conclusions}
To understand complex biological processes such as protein-ligand binding, or heterogenous catalysis, enhanced sampling methods are required to correctly modeled the process of interest and calculate the changes in the free energy that govern such reactions. In order to get results in a reasonable time and with enough accuracy, external restraints have to be incorporated into the simulations. In this work, we introduce external restraints to the family of \caps{ABF} methods incorporated into \PySAGES. These methods have the advantage of accelerate convergence by relying on neural networks or spectral decomposition. The correct free energy landscapes with the incorporation of external restraints was validated in four examples were classical and Machine Learned Force Fields were used, showing that the methods can give high accuracy with less computational time. We hope this work serves the scientific community that are applying classical and \caps{MLFF} for free energy calculations, where the flexibility and generality of external restraints are worthwhile for accurate modeling of realistic systems and phenomena.

\section{Methods}
\subsection{
  Modeling of \caps{ADP} in explicit water and coarse grained \caps{MLFF}
}
\caps{ADP} is modeled with the Amber ff99Bildn force field solvated with 1000 water molecules using the \caps{TIP3P} model. The simulation was done in the \caps{NVT} ensemble at \SI{300}{K} using the Langevin thermostat implemented in OpenMM. Spectral \caps{ABF} is used to calculate the \caps{FES} with a grid of 50 by 50. The restraint has the shape $U_{R}(\phi,\psi)=20\,(\cos(2\phi)+\sin(2\psi))$ in \si{kJ/mol} units. The Allegro coarse-grained \caps{MLFF} of \caps{ADP} is obtained from training an implicit model as depicted in~\cite{Perez-Lemus-data,depablogithub}. Spectral \caps{ABF} is used with \caps{ASE} with the langevin thermostat at \SI{300}{K} in a grid of 50 by 50. The restraint has the shape $U_{R}(\phi,\psi)=0.3125\,(\cos(2\psi)+\sin(2\phi))$ in \si{eV} units.

\subsection{Modeling of \caps{CB8--G6} system }
The \caps{CB8--G6} complex tridimensional structure was obtained from the \caps{SAMPL6} challenge repository \url{(https://github.com/samplchallenges/SAMPL6/tree/master/host_guest)}. The parametrization of the \caps{CB8} (host) and the guest ligand (\caps{G6}) was done with the general amber force field \caps{GAFF}~\cite{RM2}.
The protonation state of the guest ligand cycloheptanamine corresponded to the predicted dominant microstate at the experimental $p$H at which binding affinities were collected~\cite{rizzi2018overview}.
The charges for the complex were determined with antechamber using the \caps{AM1-BCC} method~\cite{RM3}. Approximately 5,000 \caps{TIP3P} water molecules and 1 Chloride ion were used to explicitly solvate and neutralize the complex. Prior to initiating the free energy calculations, the system underwent a minimization and equilibration protocol. The first energy minimization step included 5,000 steps, which involved 2,500 steepest descent steps where the complex was kept fix; followed by a second minimization of 8,000 steps involving 4,000 steepest descent steps. Equilibration was performed through a gradual temperature increase from \SI{0}{K} to \SI{300}{K} over \SI{500}{ps} using a Langevin thermostat with a temperature coupling constant of \SI{1}{ps} in a \caps{NVT} ensemble. Long range electrostatic interactions were considered using the Particle Mesh Ewald method with a cutoff of \SI{12}{\AA}. The collective variables used in restraint and sampling are the parallel and perpendicular projection on an axis $\xi_{\parallel},\, \xi_{\perp}$ defined by two points $\mathbf{A}$ and $\mathbf{B}$ in space,
\begin{eqnarray}
    \xi_{\parallel}&=& (\mathbf{r'}-\mathbf{A})\cdot\hat{v}\\
    \xi_{\perp}&=&\parallel(\mathbf{r}'-\mathbf{A})-\xi_{\parallel}\hat{v}\parallel\\
    \hat{v}&=& (\mathbf{B}-\mathbf{A})/\parallel \mathbf{B}-\mathbf{A}\parallel\\
    \mathbf{r}'&=&\mathcal{M}(\mathbf{r}-\mathbf{R})+\mathbf{R}_\text{ref}
\end{eqnarray}
where $\mathbf{r}$ is the position of the guest/ligand, $\mathbf{R}$ is the center of mass of the host/protein, $\mathbf{R}_\text{ref}$ is the center of mass of the reference, and $\mathcal{M}$ is the best fit rotation between the host/protein and the reference. The Kabsch algorithm was used for the rotation~\cite{lawrence2019purely}.

\subsection{Modeling of Benzamidine and Trypsin. }
The trypsin benzamidine model was obtained from the github repository of One\caps{OPES} \url{https://github.com/valeriorizzi/OneOPES}. The input gromacs files were used with the OpenMM parser with \PySAGES. The Spectral \caps{ABF} method was used with a grid of 160 points sampling from \SI{0.2}{nm} to \SI{2.5}{nm}. The \caps{RMSD} restraint constant was set to \SI{800,000}{kJ/mol.nm^2} as in the repository with the same reference PDB. The funnel shape has the form $f(\xi_{\parallel})=(Z_{0}-R_\text{cyl}) / (1+\exp{(-10(1.1-\xi_{\parallel}))}) + R_\text{cyl}$ with $Z_{0}=\SI{0.4}{nm}$, $R_\text{cyl}=\SI{0.2}{nm}$. The funnel constant was set to \SI{10,000}{kJ/mol.nm^2}. The reference free energy from~\cite{ansari2022water} was obtained from the same One\caps{OPES} repository. Three independent replicas were running starting from the gromacs input file. The \caps{NVT} simulation was set at \SI{300}{K} using the Langevin integrator with a friction constant of \SI{1}{ps^{-1}} and a timestep of \SI{1}{fs}. The nonbonded cutoff was set to \SI{1.2}{nm} with a switch distance of \SI{1}{nm}.

\subsection{Machine learned force field for methane activation.}
The optimized \caps{MLFF} for methane activation for (111) Ni surface was obtained from~\cite{xu2024molecular}. The system was simulated with \caps{ASE}\,+\,\PySAGES using the Langevin integrator at \SI{1100}{K} with a friction constant of 0.1 and a timestep of \SI{0.5}{fs}. The mass of hydrogens was increased to 2~a.u. The collective variables used are $z=\parallel \mathbf{r}_{C}-\text{\caps{COM}}_\text{Ni layer}\parallel_{z}$ with $\text{\caps{COM}}_\text{Ni layer}$ the center of mass of the top layer of Ni, and the methane coordination number
\begin{equation}
    \text{CN}_\text{C,H}=\sum_{i=1}^{4} \frac{1-(\frac{r_{ij}-d_{0}}{r_{0}})^n}{1-(\frac{r_{ij}-d_{0}}{r_{0}})^m}
\end{equation}
with $d_{0}=0$, $r_{0}=\SI{1.6}{\AA}$, $n=9$, $m=14$, and $r_{ij}$ the periodic distance between C-H atoms. The geometric restraint has the shape $f(\mathbf{z}) = ({R-1}) / ({1+\exp{(-10\,(14.5-z))}})+1.0$ with $z$ the z component of the atom measured relative to the \caps{COM} of the Ni layer, and $R=1.5,2.5,3.5$. For nickel, $k_{1}=\SI{3}{eV/\AA^2}$, for carbon and hydrogen $k_{1}=k_{2}=\SI{3}{eV/\AA^2}$, and $z_{\max}=19$~\AA\ to avoid that hydrogens go to the opposite side of the Ni layer, the simulation box height has a dimension of \SI{22.1}{\AA}. 

\section{Acknowledgments}
This work is supported by the Department of Energy, Basic Energy Sciences, Materials Science and Engineering Division, through the Midwest Integrated Center for Computational Materials (\caps{MICC}o\caps{M}).

\bibliographystyle{unsrt}
\bibliography{sample}

\begin{thebibliography}{10}

\bibitem{prot-lig}
Michael~K Gilson and Huan-Xiang Zhou.
\newblock Calculation of protein-ligand binding affinities.
\newblock {\em Annu. Rev. Biophys. Biomol. Struct.}, 36:21--42, 2007.

\bibitem{piccini2022ab}
GiovanniMaria Piccini, Mal-Soon Lee, Simuck~F Yuk, Difan Zhang, Greg Collinge,
  Loukas Kollias, Manh-Thuong Nguyen, Vassiliki-Alexandra Glezakou, and Roger
  Rousseau.
\newblock Ab initio molecular dynamics with enhanced sampling in heterogeneous
  catalysis.
\newblock {\em Catalysis Science \& Technology}, 12(1):12--37, 2022.

\bibitem{ramezani2017understanding}
Hadi Ramezani-Dakhel, Monirosadat Sadati, Mohammad Rahimi, Abelardo
  Ramirez-Hernandez, Beno{\^\i}t Roux, and Juan~J de~Pablo.
\newblock Understanding atomic-scale behavior of liquid crystals at aqueous
  interfaces.
\newblock {\em Journal of chemical theory and computation}, 13(1):237--244,
  2017.

\bibitem{burke2013dft}
Kieron Burke and Lucas~O Wagner.
\newblock Dft in a nutshell.
\newblock {\em International Journal of Quantum Chemistry}, 113(2):96--101,
  2013.

\bibitem{brooks2021classical}
Charles~L Brooks, David~A Case, Steve Plimpton, Beno{\^\i}t Roux, David Van~der
  Spoel, and Emad Tajkhorshid.
\newblock Classical molecular dynamics.
\newblock {\em The Journal of chemical physics}, 154(10), 2021.

\bibitem{unke2021machine}
Oliver~T Unke, Stefan Chmiela, Huziel~E Sauceda, Michael Gastegger, Igor
  Poltavsky, Kristof~T Sch{\"u}tt, Alexandre Tkatchenko, and Klaus-Robert
  M{\"u}ller.
\newblock Machine learning force fields.
\newblock {\em Chemical Reviews}, 121(16):10142--10186, 2021.

\bibitem{henin2022enhanced}
J{\'e}r{\^o}me H{\'e}nin, Tony Leli{\`e}vre, Michael~R Shirts, Omar Valsson,
  and Lucie Delemotte.
\newblock Enhanced sampling methods for molecular dynamics simulations.
\newblock {\em arXiv preprint arXiv:2202.04164}, 2022.

\bibitem{wtmetad}
Alessandro Barducci, Giovanni Bussi, and Michele Parrinello.
\newblock Well-tempered metadynamics: A smoothly converging and tunable
  free-energy method.
\newblock {\em Phys. Rev. Lett.}, 100:020603, Jan 2008.

\bibitem{invernizzi2020unified}
Michele Invernizzi, Pablo~M Piaggi, and Michele Parrinello.
\newblock Unified approach to enhanced sampling.
\newblock {\em Physical Review X}, 10(4):041034, 2020.

\bibitem{torrie1977nonphysical}
Glenn~M Torrie and John~P Valleau.
\newblock Nonphysical sampling distributions in monte carlo free-energy
  estimation: Umbrella sampling.
\newblock {\em Journal of computational physics}, 23(2):187--199, 1977.

\bibitem{comer2015adaptive}
Jeffrey Comer, James~C Gumbart, J{\'e}r{\^o}me H{\'e}nin, Tony Leli{\`e}vre,
  Andrew Pohorille, and Christophe Chipot.
\newblock The adaptive biasing force method: Everything you always wanted to
  know but were afraid to ask.
\newblock {\em The Journal of Physical Chemistry B}, 119(3):1129--1151, 2015.

\bibitem{fu2016extended}
Haohao Fu, Xueguang Shao, Christophe Chipot, and Wensheng Cai.
\newblock Extended adaptive biasing force algorithm. an on-the-fly
  implementation for accurate free-energy calculations.
\newblock {\em Journal of chemical theory and computation}, 12(8):3506--3513,
  2016.

\bibitem{decherchi2020thermodynamics}
Sergio Decherchi and Andrea Cavalli.
\newblock Thermodynamics and kinetics of drug-target binding by molecular
  simulation.
\newblock {\em Chemical Reviews}, 120(23):12788--12833, 2020.

\bibitem{fu2022accurate}
Haohao Fu, Haochuan Chen, Marharyta Blazhynska, Emma Goulard Coderc~de Lacam,
  Florence Szczepaniak, Anna Pavlova, Xueguang Shao, James~C Gumbart,
  Fran{\c{c}}ois Dehez, Beno{\^\i}t Roux, et~al.
\newblock Accurate determination of protein: ligand standard binding free
  energies from molecular dynamics simulations.
\newblock {\em Nature protocols}, 17(4):1114--1141, 2022.

\bibitem{woo2005calculation}
Hyung-June Woo and Beno{\^\i}t Roux.
\newblock Calculation of absolute protein--ligand binding free energy from
  computer simulations.
\newblock {\em Proceedings of the National Academy of Sciences},
  102(19):6825--6830, 2005.

\bibitem{limongelli2013funnel}
Vittorio Limongelli, Massimiliano Bonomi, and Michele Parrinello.
\newblock Funnel metadynamics as accurate binding free-energy method.
\newblock {\em Proceedings of the National Academy of Sciences},
  110(16):6358--6363, 2013.

\bibitem{chmiela2023accurate}
Stefan Chmiela, Valentin Vassilev-Galindo, Oliver~T Unke, Adil Kabylda,
  Huziel~E Sauceda, Alexandre Tkatchenko, and Klaus-Robert M{\"u}ller.
\newblock Accurate global machine learning force fields for molecules with
  hundreds of atoms.
\newblock {\em Science Advances}, 9(2):eadf0873, 2023.

\bibitem{deepmd}
Jinzhe Zeng, Duo Zhang, Denghui Lu, Pinghui Mo, Zeyu Li, Yixiao Chen,
  Mari{\'a}n Rynik, Li’ang Huang, Ziyao Li, Shaochen Shi, et~al.
\newblock {DeePMD}-kit v2: A software package for deep potential models.
\newblock {\em The Journal of Chemical Physics}, 159(5), 2023.

\bibitem{allegro}
Albert Musaelian, Simon Batzner, Anders Johansson, Lixin Sun, Cameron~J Owen,
  Mordechai Kornbluth, and Boris Kozinsky.
\newblock Learning local equivariant representations for large-scale atomistic
  dynamics.
\newblock {\em Nature Communications}, 14(1):579, 2023.

\bibitem{gissinger2017modeling}
Jacob~R Gissinger, Benjamin~D Jensen, and Kristopher~E Wise.
\newblock Modeling chemical reactions in classical molecular dynamics
  simulations.
\newblock {\em Polymer}, 128:211--217, 2017.

\bibitem{nitrogen}
Luigi Bonati, Daniela Polino, Cristina Pizzolitto, Pierdomenico Biasi, Rene
  Eckert, Stephan Reitmeier, Robert Schl{\"o}gl, and Michele Parrinello.
\newblock The role of dynamics in heterogeneous catalysis: Surface diffusivity
  and n2 decomposition on fe (111).
\newblock {\em Proceedings of the National Academy of Sciences},
  120(50):e2313023120, 2023.

\bibitem{zubieta2024pysages}
Pablo~F Zubieta~Rico, Ludwig Schneider, Gustavo~R P{\'e}rez-Lemus, Riccardo
  Alessandri, Siva Dasetty, Trung~D Nguyen, Cintia~A Men{\'e}ndez, Yiheng Wu,
  Yezhi Jin, Yinan Xu, et~al.
\newblock {PySAGES}: Flexible, advanced sampling methods accelerated with
  {GPUs}.
\newblock {\em npj Computational Materials}, 10(1):35, 2024.

\bibitem{guo2018adaptive}
Ashley~Z Guo, Emre Sevgen, Hythem Sidky, Jonathan~K Whitmer, Jeffrey~A Hubbell,
  and Juan~J De~Pablo.
\newblock Adaptive enhanced sampling by force-biasing using neural networks.
\newblock {\em The Journal of chemical physics}, 148(13), 2018.

\bibitem{sevgen2020combined}
Emre Sevgen, Ashley~Z Guo, Hythem Sidky, Jonathan~K Whitmer, and Juan~J
  De~Pablo.
\newblock Combined force-frequency sampling for simulation of systems having
  rugged free energy landscapes.
\newblock {\em Journal of chemical theory and computation}, 16(3):1448--1455,
  2020.

\bibitem{rico2025}
Pablo~F. Zubieta~Rico, Gustavo~R. Pérez-Lemus, and Juan~J. de~Pablo.
\newblock Efficient sampling of free energy landscapes with functions in
  {Sobolev} spaces.
\newblock {\em The Journal of Chemical Physics}, 162(8):084109, 02 2025.

\bibitem{darve2008adaptive}
Eric Darve, David Rodr{\'\i}guez-G{\'o}mez, and Andrew Pohorille.
\newblock Adaptive biasing force method for scalar and vector free energy
  calculations.
\newblock {\em The Journal of chemical physics}, 128(14), 2008.

\bibitem{Perez-Lemus-data}
Gustavo Perez-Lemus, Yinan Xu, Yezhi Jin, Pablo Zubieta~Rico, and Juan
  de~Pablo.
\newblock The importance of sampling the dynamical modes: Reevaluating
  benchmarks for invariant and equivariant features of machine learning
  potentials for simulation of free energy landscapes.
\newblock {\em The Journal of Chemical Physics}, 161(24):244703, 12 2024.

\bibitem{CB8}
Jay~L Banks, Hege~S Beard, Yixiang Cao, Art~E Cho, Wolfgang Damm, Ramy Farid,
  Anthony~K Felts, Thomas~A Halgren, Daniel~T Mainz, Jon~R Maple, et~al.
\newblock Integrated modeling program, applied chemical theory (impact).
\newblock {\em Journal of computational chemistry}, 26(16):1752--1780, 2005.

\bibitem{rizzi2018overview}
Andrea Rizzi, Steven Murkli, John~N McNeill, Wei Yao, Matthew Sullivan,
  Michael~K Gilson, Michael~W Chiu, Lyle Isaacs, Bruce~C Gibb, David~L Mobley,
  et~al.
\newblock Overview of the sampl6 host--guest binding affinity prediction
  challenge.
\newblock {\em Journal of computer-aided molecular design}, 32(10):937--963,
  2018.

\bibitem{bertazzo2021machine}
Martina Bertazzo, Dorothea Gobbo, Sergio Decherchi, and Andrea Cavalli.
\newblock Machine learning and enhanced sampling simulations for computing the
  potential of mean force and standard binding free energy.
\newblock {\em Journal of chemical theory and computation}, 17(8):5287--5300,
  2021.

\bibitem{roux2008comment}
Beno{\^\i}t Roux, Olaf~S Andersen, and Toby~W Allen.
\newblock Comment on “free energy simulations of single and double ion
  occupancy in gramicidin a”[j. chem. phys. 126, 105103 (2007)].
\newblock {\em The Journal of chemical physics}, 128(22), 2008.

\bibitem{doudou2009standard}
Slimane Doudou, Neil~A Burton, and Richard~H Henchman.
\newblock Standard free energy of binding from a one-dimensional potential of
  mean force.
\newblock {\em Journal of chemical theory and computation}, 5(4):909--918,
  2009.

\bibitem{HUMP96}
William Humphrey, Andrew Dalke, and Klaus Schulten.
\newblock {VMD} -- {V}isual {M}olecular {D}ynamics.
\newblock {\em Journal of Molecular Graphics}, 14:33--38, 1996.

\bibitem{ansari2022water}
Narjes Ansari, Valerio Rizzi, and Michele Parrinello.
\newblock Water regulates the residence time of benzamidine in trypsin.
\newblock {\em Nature Communications}, 13(1):5438, 2022.

\bibitem{rizzi2023oneopes}
Valerio Rizzi, Simone Aureli, Narjes Ansari, and Francesco~Luigi Gervasio.
\newblock {OneOPES}, a combined enhanced sampling method to rule them all.
\newblock {\em Journal of Chemical Theory and Computation}, 19(17):5731--5742,
  2023.

\bibitem{xu2024molecular}
Yinan Xu, Yezhi Jin, Jireh~S Garc{\'i}a~S{\'a}nchez, Gustavo~R P{\'e}rez-Lemus,
  Pablo~F Zubieta~Rico, Massimiliano Delferro, and Juan~J de~Pablo.
\newblock A molecular view of methane activation on {Ni}~(111) through enhanced
  sampling and machine learning.
\newblock {\em The Journal of Physical Chemistry Letters}, 15:9852--9862, 2024.

\bibitem{depablogithub}
\url{https://github.com/depablogroup/ConstraintFreeMLIP}.

\bibitem{RM2}
Junmei Wang, Romain~M Wolf, James~W Caldwell, Peter~A Kollman, and David~A
  Case.
\newblock Development and testing of a general amber force field.
\newblock {\em Journal of computational chemistry}, 25(9):1157--1174, 2004.

\bibitem{RM3}
Junmei Wang, Wei Wang, Peter~A Kollman, and David~A Case.
\newblock Automatic atom type and bond type perception in molecular mechanical
  calculations.
\newblock {\em Journal of molecular graphics and modelling}, 25(2):247--260,
  2006.

\bibitem{lawrence2019purely}
Jim Lawrence, Javier Bernal, and Christoph Witzgall.
\newblock A purely algebraic justification of the kabsch-umeyama algorithm.
\newblock {\em Journal of research of the National Institute of Standards and
  Technology}, 124:1, 2019.

\end{thebibliography}

\end{document}